\documentstyle[12pt]{article}

\catcode`\@=11
\long\def\@makefntext#1{
\protect\noindent \hbox to 3.2pt {\hskip-.9pt
$^{{\ninerm\@thefnmark}}$\hfil}#1\hfill}		

\def\@makefnmark{\hbox to 0pt{$^{\@thefnmark}$\hss}}    

\def\ps@myheadings{\let\@mkboth\@gobbletwo
\def\@oddhead{\hbox{}
\rightmark\hfil\ninerm\thepage}
\def\@oddfoot{}\def\@evenhead{\ninerm\thepage\hfil
\leftmark\hbox{}}\def\@evenfoot{}
\def\sectionmark##1{}\def\subsectionmark##1{}}

\renewcommand{\thefootnote}{\fnsymbol{footnote}}

\newcounter{sectionc}\newcounter{subsectionc}\newcounter{subsubsectionc}
\renewcommand{\section}[1] {\vspace*{0.6cm}\addtocounter{sectionc}{1}
\setcounter{subsectionc}{0}\setcounter{subsubsectionc}{0}\noindent
	{\normalsize\bf\thesectionc. #1}\par\vspace*{0.4cm}}
\renewcommand{\subsection}[1] {\vspace*{0.6cm}\addtocounter{subsectionc}{1}
	\setcounter{subsubsectionc}{0}\noindent
	{\normalsize\it\thesectionc.\thesubsectionc. #1}\par\vspace*{0.4cm}}
\renewcommand{\subsubsection}[1]
{\vspace*{0.6cm}\addtocounter{subsubsectionc}{1}
	\noindent {\normalsize\rm\thesectionc.\thesubsectionc.\thesubsubsectionc.
	#1}\par\vspace*{0.4cm}}

\newcounter{appendixc}
\newcounter{subappendixc}[appendixc]
\newcounter{subsubappendixc}[subappendixc]

\renewcommand{\appendix}[1] {\vspace*{0.6cm}
        \refstepcounter{appendixc}
        \setcounter{figure}{0}
        \setcounter{table}{0}
        \setcounter{equation}{0}
        \renewcommand{\thefigure}{\Alph{appendixc}.\arabic{figure}}
        \renewcommand{\thetable}{\Alph{appendixc}.\arabic{table}}
        \renewcommand{\theappendixc}{\Alph{appendixc}}
        \renewcommand{\theequation}{\Alph{appendixc}.\arabic{equation}}
        \noindent{\bf Appendix \theappendixc #1}\par\vspace*{0.4cm}}

\def\abstracts#1{{

\centering{\begin{minipage}{12.2truecm}\footnotesize\baselineskip=12pt\noindent
	\centerline{\footnotesize ABSTRACT}\vspace*{0.3cm}
	\parindent=0pt #1
	\end{minipage}}\par}}

\renewenvironment{thebibliography}[1]
	{\begin{list}{\arabic{enumi}.}
	{\usecounter{enumi}\setlength{\parsep}{0pt}
\setlength{\leftmargin 1.25cm}{\rightmargin 0pt}
	 \setlength{\itemsep}{0pt} \settowidth
	{\labelwidth}{#1.}\sloppy}}{\end{list}}

\newcounter{itemlistc}
\newcounter{romanlistc}
\newcounter{alphlistc}
\newcounter{arabiclistc}

\newcommand{\fcaption}[1]{
        \refstepcounter{figure}
        \setbox\@tempboxa = \hbox{\footnotesize Fig.~\thefigure. #1}
        \ifdim \wd\@tempboxa > 6in
           {\begin{center}
        \parbox{6in}{\footnotesize\baselineskip=12pt Fig.~\thefigure. #1}
            \end{center}}
        \else
             {\begin{center}
             {\footnotesize Fig.~\thefigure. #1}
              \end{center}}
        \fi}

\newcommand{\tcaption}[1]{
        \refstepcounter{table}
        \setbox\@tempboxa = \hbox{\footnotesize Table~\thetable. #1}
        \ifdim \wd\@tempboxa > 6in
           {\begin{center}
        \parbox{6in}{\footnotesize\baselineskip=12pt Table~\thetable. #1}
            \end{center}}
        \else
             {\begin{center}
             {\footnotesize Table~\thetable. #1}
              \end{center}}
        \fi}

\def\@citex[#1]#2{\if@filesw\immediate\write\@auxout
	{\string\citation{#2}}\fi
\def\@citea{}\@cite{\@for\@citeb:=#2\do
	{\@citea\def\@citea{,}\@ifundefined
	{b@\@citeb}{{\bf ?}\@warning
	{Citation `\@citeb' on page \thepage \space undefined}}
	{\csname b@\@citeb\endcsname}}}{#1}}

\newif\if@cghi
\def\cite{\@cghitrue\@ifnextchar [{\@tempswatrue
	\@citex}{\@tempswafalse\@citex[]}}
\def\citelow{\@cghifalse\@ifnextchar [{\@tempswatrue
	\@citex}{\@tempswafalse\@citex[]}}
\def\@cite#1#2{{$\null^{#1}$\if@tempswa\typeout
	{IJCGA warning: optional citation argument
	ignored: `#2'} \fi}}

 1
 1
 1

\font\ninerm=cmr9

\def\b{\beta}

\def\d{\delta}

\def\f{\phi}
\def\g{\gamma}
\def\h{\eta}

\def\j{\psi}

\def\m{\mu}

\def\p{\pi}
\def\q{\theta}

\def\x{\xi}

\def\vf{\varphi}

\def\bo{{\raise-.5ex\hbox{\large$\Box$}}}               
\def\pa{\partial}                                       
\def\TH{{\raise.2ex\hbox{$\displaystyle \bigodot$}\mskip-4.7mu \llap H \;}}
\def\face{{\raise.2ex\hbox{$\displaystyle \bigodot$}\mskip-2.2mu \llap {$\ddot
        \smile$}}}                                      

\def\VEV#1{\left\langle #1\right\rangle}        
\def\abs#1{\left| #1\right|}                    
\def\leftrightarrowfill{$\mathsurround=0pt \mathord\leftarrow \mkern-6mu
        \cleaders\hbox{$\mkern-2mu \mathord- \mkern-2mu$}\hfill
        \mkern-6mu \mathord\rightarrow$}
\def\dvec#1{\vbox{\ialign{##\crcr
        \leftrightarrowfill\crcr\noalign{\kern-1pt\nointerlineskip}
        $\hfil\displaystyle{#1}\hfil$\crcr}}}           

\def\frac#1#2{{\textstyle{#1\over\vphantom2\smash{\raise.20ex
        \hbox{$\scriptstyle{#2}$}}}}}                   
\def\sfrac#1#2{{\vphantom1\smash{\lower.5ex\hbox{\small$#1$}}\over
        \vphantom1\smash{\raise.4ex\hbox{\small$#2$}}}} 
\def\bfrac#1#2{{\vphantom1\smash{\lower.5ex\hbox{$#1$}}\over
        \vphantom1\smash{\raise.3ex\hbox{$#2$}}}}       
\def\afrac#1#2{{\vphantom1\smash{\lower.5ex\hbox{$#1$}}\over#2}}    

\def\half{{\frac12}}
\def\ha{\half}

\newskip\humongous \humongous=0pt plus 1000pt minus 1000pt

\newif\ifdtup

\def\np#1#2#3{{\it Nucl.~Phys.~}{\bf B{#1}} (19{#2}) #3}

\begin{document}

\centerline{\normalsize{\bf
ON THE BRST COHOMOLOGY OF $N{=}2$ STRINGS}~\footnote{
Supported in part by the `Deutsche Forschungsgemeinschaft'} }
\vspace*{0.4cm}
\centerline{\footnotesize OLAF LECHTENFELD}
\baselineskip=13pt
\centerline{\footnotesize\it
Institut f\"ur Theoretische Physik, Universit\"at Hannover }
\baselineskip=12pt
\centerline{\footnotesize\it
D--30167 Hannover, Germany }
\centerline{\footnotesize E-mail: lechtenf@itp.uni-hannover.de}
\vspace*{0.6cm}
\abstracts{
We analyze the BRST cohomology of the critical $N{=}2$ NSR string
using chiral bosonization. Picture--changing and spectral flow is
made explicit in a holomorphic field basis. The integration of
fermionic and $U(1)$ moduli is performed and yields picture-- and
$U(1)$ ghost number--changing insertions into the string measure for
$n$-point amplitudes at arbitrary genus and $U(1)$ instanton number.}

\vspace*{0.5cm}
\normalsize\baselineskip=15pt
\setcounter{footnote}{0}
\renewcommand{\thefootnote}{\alph{footnote}}
Strings with two world-sheet supersymmetries have been around for
almost 20 years,~\cite{marcus}
with recently renewed interest.~\cite{new}
The gauge-invariant $N{=}2$ string world-sheet action is given by coupling
$N{=}2$ supergravity to two complex $N{=}2$ scalar matter multiplets
$(X^\m,\j^\m)$, $\m{=}0,1$.
Superconformal gauge fixing produces conformal ghosts ($b,c$),
complex $N{=}2$ superconformal ghosts ($\b,\g$), and real $U(1)$ ghosts
($\tilde b,\tilde c$).
The $N{=}2$ superconformal algebra is generated by the currents
$T_{\rm tot}$, $G^\pm_{\rm tot}$, and $J_{\rm tot}$.

In contrast to the $N{=}1$ string, chiral bosonization of $\j^\m$ and $\b,\g$
depends on the field basis. In the `real basis', one bosonizes real and
imaginary parts of the complex NSR fermions and their ghosts, whereas
holomorphic and antiholomorphic combinations are taken in the `holomorphic
basis'. If not indicated otherwise, we shall work in the holomorphic basis
which has the advantage of diagonalizing the $U(1)$ symmetry. The $U(1)$
charges appear as $\pm$ superscripts.  In this way,
$\j^{\pm\m}$ and $\b^\pm,\g^\pm$ are replaced by two pairs of bosons,
$\f^\pm$ and $\vf^\pm$, plus two auxiliary fermion systems ($\h^\pm,\x^\pm$),
spanning an extended Fock space containing ${\bf Z}{\times}{\bf Z}$
copies of the original fermionic one.
It is easy to see that BRST non-trivial operators must have vanishing conformal
dimension and $U(1)$ charge. Further grading of the cohomology is effected
by the mass level, the total ghost number~$u\in{\bf Z}$,
and two picture numbers~$\p^\pm\in\ha{\bf Z}$, with $\p^+{+}\p^-\in{\bf Z}$.
Integral and half-integral picture numbers correspond to NS and R states,
respectively.
For generic momenta, we find\cite{bkl} that the BRST cohomology on the massless
level consists of {\it four\/} classes of states for each pair $(\p^+,\p^-)$,
labelled by $v{\equiv}u{-}\p^+{-}\p^-\in\{1,2,2,3\}$ and created by
vertex operators of type $c$, $\tilde c$, $c\pa c$ and $\tilde c c\pa c$.

Physical states correspond to {\it classes\/} of BRST cohomology classes,
formed under the following four equivalence relations.
First, $c$-- and $c\pa c$--type vertices are to be identified just as in the
bosonic string.
Second, $\tilde c$--type vertex operators get converted to others by applying
the $U(1)$ ghost number--changing operator~$Z^0$.
Third, two picture--changing operators~$Z^\pm$ raise the picture numbers
of vertex operators by unit amounts.
Fourth, NS and R states are connected by the action of the spectral-flow
operators~$S\!F\!O^\pm$ which move $(\p^+,\p^-)\to(\p^+{\pm}\ha,\p^-{\mp}\ha)$.
These maps are given by
$$
Z^0 = \oint\!\tilde b\ \d\Bigl(\oint\! J_{\rm tot}\Bigr) \quad,\quad
Z^\pm(z) = \d(\b^\pm)\ G^\pm_{\rm tot} \quad,\quad
S\!F\!O^\pm(z) = \exp\{\pm\ha\int^z\! J_{\rm tot}\} \quad,
\eqno(1) $$
and commute with $Q_{\rm BRST}$ but are non-trivial.
In this fashion, each physical state has a representative $V^{\rm can}$ at
$v{=}2$ ($\tilde cc$--type) in the canonical picture $(\p^+,\p^-)=({-}1,{-}1)$.
On the massless level, only a single scalar excitation survives.
For the computation of string amplitudes, however, vertex operators in various
other ghost and picture sectors are useful and have been constructed.\cite{bkl}

Any $n$--point amplitude involves a sum over genera $h\in{\bf Z}^+$ and
$U(1)$ instanton number~$c\in{\bf Z}$.
To compute the contribution for fixed $h$ and~$c$,
one must integrate out $2h{-}2{\pm}c{+}n$ complex fermionic moduli of
$U(1)$ charge ${\pm}1$, respectively, and $h{-}1{+}n$ complex $U(1)$ moduli,
to obtain an integration measure for the remaining $3h{-}3{+}n$ complex
metric moduli.\footnote{
As always, the cases of the sphere and the torus require some modifications.}~
The result vanishes for $|c|{>}2h{-}2{+}n$
and symbolically reads\cite{kl}
$$
\VEV{ \abs{
(\oint\!b)^{3h-3+n}\ (Z^+)^{2h-2+c+n}\ (Z^-)^{2h-2-c+n}\ (Z^0)^{n-1} }^2
\prod_{i=1}^{h}\! \Bigl[ Z^0(a_i)Z^0(b_i) \Bigr]\
V_1^{\rm can}\!\ldots V_n^{\rm can} }
\eqno(2) $$
where $a_i$ and $b_i$ denote the homology cycles.
The picture--changers $Z^\pm$ and $Z^0$ may be used partially to
convert vertex operators to other pictures and/or ghost numbers.

Invariance of correlation functions under spectral flow follows from the
fact that a change in monodromies for the world-sheet fermions is equivalent
to a shift in the integration over $U(1)$ moduli, which are nothing but
the flat $U(1)$ connections on the $n$--punctured genus--$h$ Riemann surface.
It is realized on the vertex operators by\cite{kl}
$$
V(z)\ \longrightarrow\ V^\q(z)\ =\
\exp\Bigl\{ \q\int^z \! J_{\rm tot}(z')\,dz' \Bigr\}\ V(z) \quad,
\eqno(3) $$
with $\q{=}{\pm}\ha$ leading to $S\!F\!O^\pm$ mapping from NS to R$^\pm$
sectors.\footnote{
The two sectors R$^+$ and R$^-$ differ
by a unit change in instanton number~$c$.}~
Stated differently,
$$
\Bigl\langle V_1\,V_2\ldots V_n \Bigr\rangle\ =\
\VEV{V_1^{\q_1}\,V_2^{\q_2}\ldots V_n^{\q_n}} \qquad{\rm for}\qquad
\sum_\ell \q_\ell =0 \quad,
\eqno(4) $$
equating all $n$--point amplitudes with the same values for $h$ and~$c$.

This work was done in collaboration with S.~Ketov.
We acknowledge fruitful discussions with N.~Berkovits, H.~L\"u, H.~Ooguri,
and C.~Pope.


\begin{thebibliography}{9}
\bibitem{marcus} see, e.g., N. Markus, hep-th/9211059, and references therein.
\bibitem{new} N. Berkovits and C. Vafa, \np{433}{95}{123};
H. L\"u and C.N.~Pope, hep-th/9411101;
N. Berkovits, hep-th/9412179 and hep-th/9503099.
\bibitem{bkl} J. Bischoff, S. Ketov and O. Lechtenfeld, \np{438}{95}{373};\\
see also: A.Giveon and M. Ro\v cek, \np{400}{93}{145}.
\bibitem{kl} S. Ketov and O. Lechtenfeld, hep-th/9503232.
\end{thebibliography}
\end{document}